\documentclass[fleqn,10pt]{wlscirep}
\usepackage[utf8]{inputenc}
\usepackage[T1]{fontenc}
\title{Impact of the Central Frequency of Environment on Non-Markovian Dynamics in Piezoelectric Optomechanical Devices }

\author[1]{Quanzhen Ding}
\author[2,3]{Peng Zhao}
\author[4,+]{Yonghong Ma}
\author[2,*]{Yusui Chen}

\affil[1]{Department of Physics, and Center for Quantum Science and Engineering, Stevens Institute of Technology, Hoboken, NJ 07030, USA}
\affil[2]{Department of Physics, New York Institute of Technology, Old Westbury, NY 11568, USA}
\affil[3]{Department of Engineering and Computing Sciences, New York Institute of Technology, Old Westbury, NY 11568, USA}
\affil[4]{School of Science, Inner Mongolia University of Science and Technology, Baotou 014010, China}

\affil[*]{yusui.chen@nyit.edu}
\affil[+]{myh\_dlut@126.com}


\begin{abstract}
The piezoelectric optomechanical devices supply a promising experimental platform to realize the coherent and effective control and measurement on optical circuits working in Terahertz (THz) frequencies via superconducting electron devices typically working in Radio (MHz) frequencies. However, quantum fluctuations are unavoidable when the size of mechanical oscillators enter into the nanoscale. The consequences of the noisy  environment is still challenging due to the lack of analytical tools. In this paper, a semi-classical and full-quantum model of piezoelectric optomechanical systems coupled to a noisy bosonic quantum environment are introduced and solved in terms of quantum-state diffusion (QSD) trajectories in the non-Markovian regime. We show that the noisy environment, particularly the central frequency of environment, can enhance the entanglement generation between optical cavities and LC circuits in some parameter regimes. Moreover, we observe the critical points in the coefficient functions, which can lead the different behaviors in the system. Besides, we also witness the entanglement transfers between macroscopic objects due to the memory effect of the environment. Our work can be applied in the fields of electric/ optical switches, and  long-distance distribution in large-scale quantum network. 
\end{abstract}
\begin{document}

\flushbottom
\maketitle
%
%


\section*{Introduction}

Optomechanical systems have received substantial interests as a promising experimental platform to improve the resolution and precision of measurement to beat the standard quantum limit, to observe macroscopic quantum phenomena, and to realize sideband cooling and parametric amplification\cite{OM:aspe2014,OM:Brooks2012,OM:Mancini1998,OM:NImm2014,OMVitali2007,OM:otter2018,OM:Kippenberg1172,OM:Weis1520,Ma2017,YChenJPB,OM:Baldacci:2016aa,OM:villafa:2018,OM:Boales:2017aa,OM:Fang:2016aa,OM:Renninger:2018aa,OM:Santos:2017,OM:cripe:2018}. Its applications, ranging from quantum information to quantum sensing, have been thoroughly studied. Recently, conventional optomechanical systems consisting of the coupled mechanical mode and cavity field have been extended to piezoelectric optomechanical devices in which radio frequency (RF) and cavity field are parametrically coupled through the coupling with the mechanical oscillator\cite{POM:han:2020,POM:wu:2020,POM:stockill:2019,POM:barg:2018,POM:dahmani:2020,POM:Han:2020aa,POM:zou:2020,POM:shumeiko:2016,POM:arrangoiz:2016,POM:wang:2019,POM:zhong:2020,POM:weiss:2018,POM:yamazaki:2020,POM:Javerzac:2016,POM:Balram:2016,POM:Jiang:2020aa,POM:Midolo:2018aa,POM:Rueda:2019aa,POM:Sychev:2018aa,POM:Tavernarakis:2018aa,POM:Wang:2018aa,POM:Witmer:2017aa}. Such hybrid systems supply an experimental platform in which the THz optical field and GHz acoustic waves are combined. In addition, the mechanical oscillator can be driven by RF signals and be measured precisely by interferometers in the optical domain, which raises significant interests for classical signal processing. 

\begin{figure}[htp]
\centering
\includegraphics[width=10 cm]{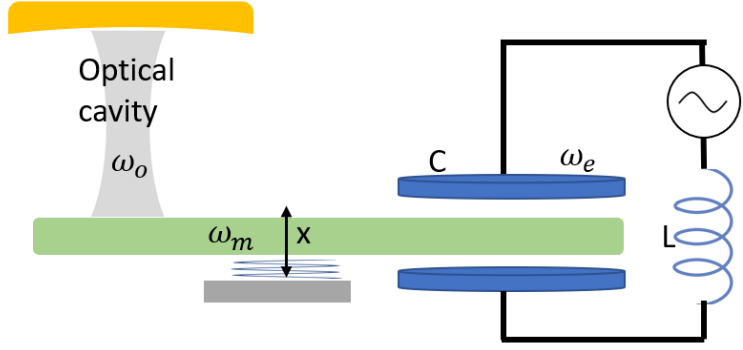}
\caption{A schematic of our model, consisting of a piezo-mechanical resonator coupled with an optical cavity and a LC circuit simultaneously.}
\label{fig:model}
\end{figure}

However, it remains a tremendous challenge to suppress the impacts of the noisy environment\cite{nielson2002,Breuer2010book}. Various schemes have been investigated to address these challenges, including eliminating the back-action noise by using the coherent quantum noise cancellation (CQNC) scheme and reducing the shot noise by lowering the input field power\cite{OM:aspe2014,YChenJPB,chen2018QIC, chen2019SCP,chen2020JPB}. Thermal noise is another source of noise unless the temperature is very low. But when the system is cooled down to few milli-Kelvin, quantum fluctuations exceed the thermal noise and become the major noise. So a fully understanding of the dynamics of the piezoelectric optomechanical (POM) system in the present of noisy environment is crucial. This paper addresses the evolution of POM systems starting from a complete microscopic Hamiltonian. Our proposed model consists of a mechanical oscillator coupled to an optical cavity and an LC circuit simultaneously, as shown in Fig.~\ref{fig:model}. 

Traditional methods to deal with our proposed model are limited in Markovian regime\cite{Breuer2010book}, assuming that the environment is too large to back-act on the system and the relaxation time of the system is much shorter comparing to the operational time. However, many future applications require high-frequency operation, for example the quantum network and quantum computing. Without loss of generalization, we need to initiate our discussion in non-Markovian regime. Moreover, the non-Markovian environment can parametrically tune the dynamics of the system and enhance the entanglement in some particular regime, unlike the Markovian environment in which only an exponential decay can be observed.

In this paper, we investigate the POM model that a mechanical oscillator, coupled to an optical cavity and an LC circuit simultaneously and embedded in a non-Markovian environment. The aim of the paper is twofold. Firstly, we derive a non-Markovian stochastic Schr\"{o}dinger equation (SSE) to characterize the dynamics of the system by using the fully quantized Hamiltonian\cite{strunz1998,chen2016book}. Each numerical solution of the SSE is so-called a trajectory and the reduced density operator of the system can be reproduced by taking average of all trajectories. Moreover, the corresponding master equation can be obtained formally\cite{chen2014,chen2014_2,chen2018QIC}. Secondly, we investigate how the entanglement is influenced by various non-Markovian environment\cite{chen2019SCP,chen2020JPB}. By adjusting the frequencies of both optical and mechanical mode, coupling strength and central frequency of the environment, we carefully study the entanglement transfer inside the POM system. We discuss the physical mechanism in the model and the experimental feasibility of our proposed model.

\section*{Background}
In this section, we briefly review the relevant background of non-Markovian dynamics, particularly using the quantum-state diffusion (QSD) approach and the corresponding master equation (MEQ) approach\cite{chen2014,chen2018QIC}. 

\subsubsection*{General non-Markovian dynamics of open quantum systems}
A general open quantum system can be studied in the system-environment framework and the formal Hamiltonian is expressed as
\begin{align*}
    H_{tot} = H_{sys} +H_{int} + H_{env}.
\end{align*}
The quantized  bosonic environment Hamiltonian, and the interaction Hamiltonian can be expressed as
\begin{align*}
    H_{env} &= \sum_k \omega_k d_k^\dagger d_k, \\
    H_{int} &= \sum_k (g_k L d_k^\dagger + g_k^*L^\dagger d_k ).
\end{align*}
where $\omega_k$ is the eigen frequency of the $k$th mode in the environment. The coupling between system and environment is characterized by the Lindblad operator $L$. Under some conditions, e.g. the environment is too large to allow the existence of back flows, and the system-environment coupling is relatively weak, the state of the system can be solely depended on the state at the last time instant. Due to the so-called Born-Markov approximation, the dynamics of the system can be described by using Lindblad master equations, Redfield master equations, quantum jump, Heisenberg approach etc. However, generalizing a non-Markovian quantum dynamics is difficult. 

The QSD approach is a method to describe the non-Markovian dynamics of the system by tuning it into a set of continuous time stochastic processes. Each of the stochastic evolution of the state is called a trajectory. And the reduced density operator of the system can be regenerated numerically as the ensemble average of all trajectories. In the interaction picture, the QSD equation is given as
\begin{align}
    \partial_{t}\psi_t = (-iH_{sys} + L z_t^* - L^\dagger \int_0^t ds\  \alpha (t,s)\frac{\delta }{\delta z_s^*} )\psi_t, 
    \label{qsd1}
\end{align}
where $\psi_t = \langle z|\Psi_{tot} \rangle$ is the continuous time stochastic trajectory of the state of the system. $z_t^* = -i \sum_k g_k z_k^* e^{-i\omega_k t}$ is a complex Gaussian stochastic process. $|z\rangle = \otimes_k |z_k\rangle$ is the collective basis of all modes in the environment, where the $|z_k\rangle$ is the Bargamann coherent state of the $k$th mode, defined as $d_k|z_k\rangle = z_k|z_k\rangle$. With the completeness identity $I = \int \frac{d^2z}{\pi} e^{-|z|^2} |z\rangle \langle z| $, the reduced density operator can formally expressed as
\begin{align}
    \rho_{sys} = \mathcal{M} [|\psi_t\rangle\langle\psi_t|] = \int_0^t \frac{d^2z}{\pi}e^{-|z|^2} |\psi_t\rangle\langle\psi_t|,
    \label{ensembleavg}
\end{align}
where the symbol $\mathcal{M}[\cdot]$ represents the ensemble average. Moreover, Gaussian stochastic processes satisfy that $\mathcal{M}[z_t] =0$, and are characterized by the correlation functions:  $\mathcal{M}[z_tz_s] = 0$, and $\mathcal{M}[z_t^*z_s] = \alpha (t,s)$. In particular, $\alpha(t,s)$ is defined as 
\begin{align}
    \alpha(t-s)= \int d\omega \ J(\omega) e^{-i\omega (t-s)} ,
    \label{corr}
\end{align}
and depends solely on the spectral function $J(\omega)=\sum_k|g_k|^2 \delta(\omega -\omega_k)$, which describes how the system-environment coupling strength is frequency depended. For instance, the correlation function $\alpha(t,s)$ will take a form of Dirac delta function when the distribution of $J(\omega)$ is flat. By considering different coupling spectral functions and the corresponding correlation functions, the QSD approach can be applied to study a variety of types of environments. 

After replacing the functional derivative $\frac{\delta \psi_t}{\delta z_s^*}$ by the product of an operator $O(t,s,z^*)$ in the system's Hilbert space and the stochastic wave function, $\frac{\delta \psi_t}{\delta z_s^*}\ = O(t,s,z^*)\psi_t$, the time-local QSD equation  (\ref{qsd1}) can be transformed to
\begin{align}
    \partial_{t}\psi_t = (-iH_{sys} + L z_t^* - L^\dagger \int_0^t ds \  \alpha (t,s)O(t,s) )\psi_t, 
    \label{qsd_eq}
\end{align}
where the $O$ operator can be determined by an evolution equation that
\begin{align}
    \partial_{t}O &=\left[-iH_{sys}+Lz_{t}^{*}-L^{\dagger}\bar{O},O\right]-L^{\dagger}\frac{\delta\bar{O}}{\delta z_{s}^{*}},
    \label{Oeq}
\end{align}
together with its initial condition that $O(t,s=t)= L$, where $\bar{O} = \int_0^t ds \alpha(t,s) O(t,s)$. 

With the QSD equation, the corresponding non-Markovian master equations can be derived as
\begin{align}
    \partial_t \rho_{sys} = -i[H_{sys},\:\rho_{sys}] +[L,\:\mathcal{M}(|\psi_t\rangle\langle\psi_t|\bar{O}^\dagger)] - [L^\dagger,\:\mathcal{M}(\bar{O}|\psi_t\rangle\langle\psi_t|)].
    \label{fmeq}
\end{align}

\subsubsection*{Macroscopic entanglement}
To study the entanglement dynamics between components in the system, we employ the negativity to measure the scale of entanglement,
\begin{align*}
    \mathcal{N}(\rho) =\frac{||\rho^{\Gamma_A}||_1 -1}{2}= \sum_i \frac{|\lambda_i|-\lambda_i}{2},
\end{align*}
where $\rho^{\Gamma_A}$ is the partial transpose of $\rho$ with respect to the subsystem $A$. $||\cdot||_1$ represents the trace norm. And $\{\lambda_i\}$ are all negative eigenvalues of $\rho^{\Gamma_A}$. The quantity $\mathcal{N}(\rho)$ is monotonic to the upper boundary of the distillable entanglement.

\section*{The Model}
The model we proposed is illustrated in Fig. 1. An optomechanical system, consisting of an optical cavity and a movable mirror, is coupled to a high-frequency piezo-mechanical resonator. As a result, the mechanical oscillator is coupled to the optical cavity and the capacitor of the superconducting LC resonator simultaneously, and achieves the coupling between the optical field and the microwave field. Moreover, a quantum environment coupled to the mechanical oscillator may be introduced by a variety of imperfections, e.g., due to diffraction, external fields, lattice vibrations etc. Thus, the system Hamiltonian can be expressed as (setting $\hbar=1$)
\begin{align}
H_{sys} & =-\omega_{o}a^{\dagger}a+\omega_{m}b^{\dagger}b+\omega_{e}c^{\dagger}c -g_{om}a^\dagger a\hat{x}-g_{me} \hat{x}\left( c+ c^{\dagger}\right)
\end{align}
where $a(a^\dagger)$, $b(b^\dagger)$, and $c(c^\dagger)$ are the annihilation (creation) operators of the optical field, the mechanical mode, and the L-C resonator respectively, with the eigen frequency of $\omega_o$, $\omega_m$ and $\omega_e$. The displacement operator $\hat{x}= x_{ZPF}(b+b^\dagger)$, where $x_{ZPF} = \sqrt{1/2m\omega_m}$ is the zero point frequency of the mechanical mode. The optomechanical interaction Hamiltonian $g_{om}\hat{x}a^\dagger a$ is nonlinear for the displacement of the movable mirror $\hat{x}$, which only is valid when the mechanical oscillation frequency is relatively smaller than the free spectral range of the cavity. The piezoelectric interaction Hamiltonian $g_{me}\hat{x}(c+ c^\dagger)$ describes a linear coupling, where the coupling strength depends on the coupling between the strain and the electric field. 

It is difficult to analytically solve the dynamics of the system, particularly when the optomechanical coupling in nonlinear. However, in the semi-classical limit, the optical mode operator $a$ can be linearized and expended as the sum of the average amplitude $\alpha$ and the fluctuation $
\delta a$, that $a=\alpha + \delta a$. As a result,the nonlinear interaction $\hat{x} a^\dagger a$ can be rewritten as $\hat{x} \left(\alpha^* + \delta a^\dagger\right)\left( \alpha + \delta a\right)$. Assuming that the fluctuation $\delta a$ is much smaller than the average amplitude $\alpha$, the final linearization of the interaction Hamiltonian can be expressed as
\begin{align*}
    H_{om} = g_{om}\hat{x}(a^\dagger + a).
\end{align*}

\subsubsection*{System-Environment couplings}
To consider a dissipative environment, the interaction Hamiltonian is given
\begin{align*}
    H_{int} = b\sum_k g_k d_k^\dagger + b^\dagger\sum_k g_k^* d_k.
\end{align*}
There are some typical spectral functions such as Lorentzian, Ohmic, sub-Ohmic, and super-Ohmic for the structure of the environment. In order to observe how the dynamics of the system gradually transits from the Markovian regime to the non-Markovian regime, we consider the Lorentzian spectral density as
\begin{align*}
    J(\omega) = \frac{1}{2\pi}\frac{\Gamma \gamma^2}{(\omega - \Omega)^2+\gamma^2},
\end{align*}
where $\Omega$ is the central frequency of the environment. The parameter $\gamma$ defines the spectral width of the coupling and $1/\gamma$ indicates the memory time of the environment. $\Gamma$ is the global coupling strength. By substituting $J(\omega)$ into Eq. (\ref{corr}), the correlation function $\alpha(t,s)$ can be
determined analytically, in the form of the Ornstein–Uhlenbeck $(O-U)$ correlation function,
\begin{align*}
    \alpha(t,s)= \frac{\Gamma \gamma}{2}e^{-\gamma|t-s| -i\Omega(t-s)}.
\end{align*}

The analytical solution of the dissipative dynamics of the system depends on the system Hamiltonian and the spectral density of the environment simultaneously. In this section, we introduce two classes of coupling in the system, and in each case, we study the non-Markovian effects influenced by the central frequency of the environment.

\subsubsection*{Case I: Weak coupling}
For the piezoelectric interaction, if the coupling strength $g_{me}$ is weak, comparing to the eigen frequencies of the mechanical mode and the LC resonator $\omega_{m}$ and $\omega_{e}$, then it is reasonable to use the rotating-wave approximation to simplify
\begin{align*}
    H_{me}=g_{me}(b+b^\dagger)(c+c^\dagger) \approx g_{me}(bc^\dagger+ b^\dagger c).
\end{align*}

And consequently, the interaction of the optical field and the mechanical oscillator can approximate to
\begin{align*}
    g_{om}\left(a+a^\dagger\right)\left(b+b^\dagger\right) \approx g_{om}\left(ab^\dagger + a^\dagger b\right).
\end{align*}

From Eq. (\ref{Oeq}), the ansatz of $O$ operator must take the form of 
\begin{align*}
O\left(t,s\right) =f_{1}\left(t,s\right)b+f_{2}\left(t,s\right)a+f_{3}\left(t,s\right)c,
\end{align*}
wither three to-be determined coefficient functions $f_1(t,s)$, $f_2(t,s)$, and $f_3(t,s)$. Substituting the ansatz into Eq. (\ref{Oeq}), the three coefficient functions can be determined by a group of partial differential equations
\begin{align*}
\partial_{t}f_{1} & =i\omega_{m}f_{1}-ig_{om}f_{2}-ig_{me}f_{3}+F_{1}f_{1},\\
\partial_{t}f_{2} & =-i\Delta_{o}f_{2}-ig_{om}f_{1}+F_{2}f_{1},\\
\partial_{t}f_{3} & =i\omega_{e}f_{1}-ig_{me}f_{1}+F_{3}f_{1},
\end{align*}
with the intial conditions
\begin{align*}
f_{1}\left(t,t\right) & =1,\\
f_{2}\left(t,t\right) & =f_{3}\left(t,t\right)=0.
\end{align*}

Due to the exponential $O-U$ style correlation function $\alpha\left(t,s\right)=\frac{\Gamma\gamma}{2}e^{-\gamma\left|t-s\right|-i\Omega\left(t-s\right)} $, we can continue to analytically derive the evolution function of the operator $\bar{O} = F_{1}\left(t\right)b+F_{2}\left(t\right)a+F_{3}\left(t\right)c$, where $F_j(t) = \int_0^t ds \alpha(t,s) f_j(t,s)$. Hence the $\bar{O}$ operator can be numerically found by the following group of expressions,
\begin{align*}
\partial_{t}F_{1} & =\frac{\Gamma\gamma}{2}-\left(\gamma+i\Omega-i\omega_{m}\right)F_{1}-ig_{om}F_{2}-ig_{me}F_{3}+F_{1}^{2},\\
\partial_{t}F_{2} & =-\left(\gamma+i\Omega+i\Delta_{o}\right)F_{2}-ig_{om}F_{1}+F_{1}F_{2},\\
\partial_{t}F_{3} & =-\left(\gamma+i\Omega-i\omega_{e}\right)F_{3}-ig_{me}F_{1}+F_{1}F_{3}.
\end{align*}

\begin{figure}[htp]
\centering
\includegraphics[width=10 cm]{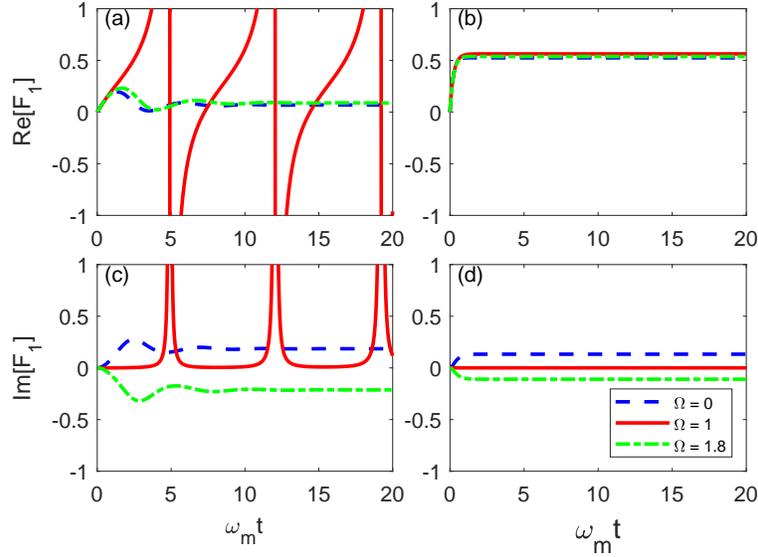}
\caption{Evolution of the real ($a, b$) and imaginary ($c, d$) value of the complex coefficient function $F_1(t)$ in the $\bar{O}$ operator for Case I.  In (a) and (c), $\gamma = 0.5$. In (b) and (d), $\gamma = 5.0$.  Other parameters are chosen as: $\omega_o = \omega_e =\omega_m$, $g_{om} = g_{me} = 0.1 \omega_m$, $\Gamma =1$. }
\label{fig:M1_F1}
\end{figure}

In Fig. \ref{fig:M1_F1}, the dynamics of the coefficient $F_1(t)$ in the $\bar{O}$ operator is illustrated when the central frequency of environment $\Omega$ equals $0, 1$ and $1.6$. We only plot $F_1$ as it is one order of magnitude larger than $F_2$ and $F_3$, and takes the dominant role in the $\bar{O}$ operator. In Fig. \ref{fig:M1_F1} (a, c), $F_1$'s real and imaginary values oscillate in the short time interval and gradually approach to a stable value in the long-time limit for $\Omega = 0$ (blue dashed line), $1.8$ (green dash-dotted line). When $\Omega \approx 1$ (red solid line), singularities emerge in the real and imaginary parts of $F_1(t)$, which also show a periodic behavior. In Fig.~\ref{fig:M1_F1}(b, d), when the decay factor $\gamma = 5$, $F_1$ approaches to the stable value $\frac{\Gamma}{2}$ quickly. In addition, since the $\bar{O}$ operator is noise free, thus $\mathcal{M}[|\psi_t\rangle\langle\psi_t|\bar{O}^\dagger] = \rho_{sys}\bar{O}^\dagger$ and consequently the formal master equation can be explicitly written as 
\begin{align*}
\partial_{t}\rho_{sys} & =-i\left[H_{sys},\rho_{sys}\right]+\left[L,\rho_{sys}\bar{O}^{\dagger}\right] - \left[L^\dagger,\:\bar{O}\rho_{sys}\right]\\
 & =-i\left[-\omega_{o}a^{\dagger}a+\omega_{m}b^{\dagger}b+\omega_{e}c^{\dagger}c-g_{om}\left(ab^{\dagger}+a^{\dagger}b\right)-g_{me}\left(bc^{\dagger}+b^{\dagger}c\right),\rho_{sys}\right]\\
 & \quad+\left[b,\rho\left(F_{1}^{*}b^{\dagger}+F_{2}^{*}a^{\dagger}+F_{3}^{*}c^{\dagger}\right)\right]+\left[\left(F_{1}b+F_{2}a+F_{3}c\right)\rho_{sys},b^{\dagger}\right].
\end{align*}

\subsubsection*{Case II: Strong coupling}
In this section, we will consider a more general coupling that the mechanical oscillator is strongly coupled to the optical cavity and the LC circuit, therefore the system Hamiltonian is
\begin{align*}
   H_{sys} & =-\omega_{o}a^{\dagger}a+\omega_{m}b^{\dagger}b+\omega_{e}c^{\dagger}c -g_{om}\left(a^\dagger +a\right)\hat{x}-g_{me} \hat{x}\left( c+ c^{\dagger}\right).
\end{align*}

\begin{figure}[htp]
\centering
\includegraphics[width=10 cm]{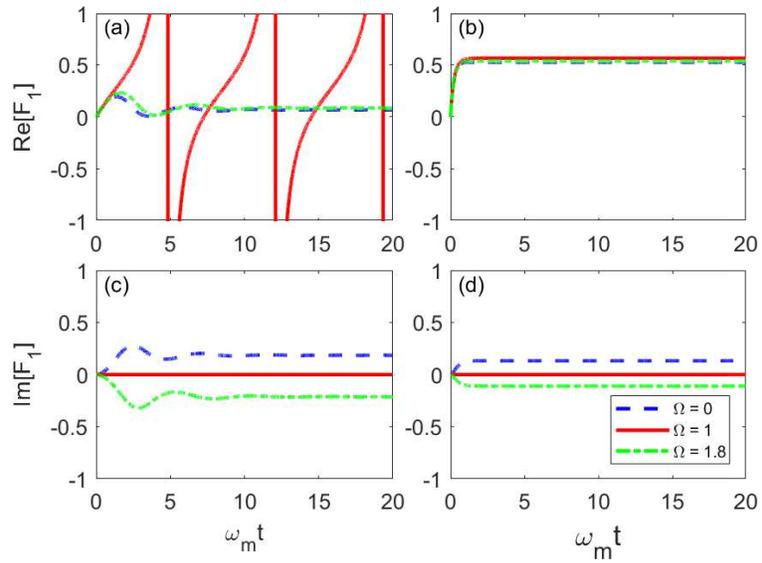}
 \caption{Evolution of the real and imaginary value of the complex coefficient function $F_1(t)$. In (a) and (c), $\gamma = 0.5$, non-Markovian regime. In (b) and (d), $\gamma = 5.0$, Markovian regime. Parameters are chosen as: $\omega_o = \omega_e =\omega_m$, $g_{om} = g_{me} =0.1 \omega_m$, $\Gamma =1$. }
\label{fig:M2_F1}
\end{figure}

The anstz of $O$ operator satisfying Eq. (\ref{Oeq}) may be found  
\begin{align*}
O(t,s) &= O_0(t,s) + \int_{0}^{t}ds'O_1(t,s,s') z_{s'},\\
O_{0}\left(t,s\right) & =f_{1}\left(t,s\right)b+f_{2}\left(t,s\right)b^{\dagger}+f_{3}\left(t,s\right)a+f_{4}\left(t,s\right)a^{\dagger}+f_{5}\left(t,s\right)c+f_{6}\left(t,s\right)c^{\dagger},\\
O_{1}\left(t,s,s'\right) &=f_{7}\left(t,s,s'\right)I,
\end{align*}
where $O_{0}(t,s)$ and $O_1(t,s,s')$ represent the noise-free and first order noise terms in the $O$ operator respectively. Similarly, by substituting the anstz of the $O$ operator into Eq. \ref{Oeq}, all coefficient functions can be numerically found from the following expressions
\begin{align*}
\partial_{t}f_{1} & =i\omega_{m}f_{1}-ig_{om}f_{3}+ig_{om}f_{4}-ig_{me}f_{5}+ig_{me}f_{6}+F_{1}f_{1},\\
\partial_{t}f_{2} & =-i\omega_{m}f_{2}-ig_{om}f_{3}+ig_{om}f_{4}-ig_{me}f_{5}+ig_{me}f_{6}+2F_{2}f_{1}-F_{1}f_{2}+F_{4}f_{3}-f_{3}F_{4}+F_{6}f_{5}-F_{5}f_{6}-F_{7}\left(t,s\right),\\
\partial_{t}f_{3} & =-i\Delta_{o}f_{3}-ig_{om}f_{1}+ig_{om}f_{2}+F_{3}f_{1},\\
\partial_{t}f_{4} & =i\Delta_{o}f_{4}-ig_{om}f_{1}+ig_{om}f_{2}+F_{4}f_{1},\\
\partial_{t}f_{5} & =i\omega_{e}f_{5}-ig_{me}f_{1}+ig_{me}f_{2}+F_{5}f_{1},\\
\partial_{t}f_{6} & =-i\omega_{e}f_{6}-ig_{me}f_{1}+ig_{me}f_{2}+F_{6}f_{1},\\
\partial_{t}f_{7} & =F_{7}\left(t,s'\right)f_{1},
\end{align*}
with the initial and boundary conditions
\begin{align*}
f_{1}\left(t,t\right) & =1,\\
f_{2}\left(t,t\right) &= f_{3}\left(t,t\right) =f_{4}\left(t,t\right)=f_{5}\left(t,t\right)=f_{6}\left(t,t\right)=0,\\
f_{7}\left(t,s,s'=t\right) & =f_{2}(t,s),\\
f_{7}\left(t,s=t,s\right) & =0.
\end{align*}
Consequently, the $\bar{O}$ may be determined by
\begin{align*}
\partial_{t}F_{1} & =\frac{\Gamma\gamma}{2}-\left(\gamma+i\Omega-i\omega_{m}\right)F_{1}-ig_{om}F_{3}+ig_{om}F_{4}-ig_{me}F_{5}+ig_{me}F_{6}+F_{1}^{2},\\
\partial_{t}F_{2} & =-\left(\gamma+i\Omega+i\omega_{m}\right)F_{2}-ig_{om}F_{3}+ig_{om}F_{4}-ig_{me}F_{5}+ig_{me}F_{6}+F_{1}F_{2}-\tilde{F}_{7},\\
\partial_{t}F_{3} & =-\left(\gamma+i\Omega+i\Delta_{o}\right)F_{3}-ig_{om}F_{1}+ig_{om}F_{2}+F_{1}F_{3},\\
\partial_{t}F_{4} & =-\left(\gamma+i\Omega-i\Delta_{o}\right)F_{4}-ig_{om}F_{1}+ig_{om}F_{2}+F_{1}F_{4},\\
\partial_{t}F_{5} & =-\left(\gamma+i\Omega-i\omega_{e}\right)F_{5}-ig_{me}F_{1}+ig_{me}F_{2}+F_{1}F_{5},\\
\partial_{t}F_{6} & =-\left(\gamma+i\Omega+i\omega_{e}\right)F_{6}-ig_{me}F_{1}+ig_{me}F_{2}+F_{1}F_{6},
\end{align*}
and
\begin{align*}
\partial_{t}\tilde{F}_{7}  =\frac{\Gamma\gamma}{2}F_{2}-2\left(\gamma+i\Omega\right)\tilde{F}_{7}+F_{1}\tilde{F}_{7},
\end{align*}
where $\tilde{F}_7(t) = \int_0^t ds \alpha(t,s)F_7(t,s)$. Numerical estimations for $F_1(t)$ and $F_2(t)$ are illustrated in Fig. \ref{fig:M2_F1}. Singularities are observed in Fig. \ref{fig:M2_F1} (a,c), the non-Markovian evolution of both $F_1$ and $F_2$, when $\Omega$ is close to $1.18$. Although the exact $O$ operator is obtained, the master equation is still hard to derive, as the $O$ operator consists of the first order noise. According to a systematic method that derives master equations\cite{chen2014}, the ensemble average $\mathcal{M}\left[P_{t}\bar{O}^{\dagger}\right]$ can be derived analytically as,
\begin{align*}
R\left(t\right) =\mathcal{M}\left[P_{t}\bar{O}^{\dagger}\right]
=\rho_{t}\bar{O}_{0}^{\dagger}+\iint_{0}^{t}ds_{1}ds_{2}\alpha\left(s_{1},s_{2}\right)O_{0}\left(t,s_{2}\right)\rho_{t}\bar{O}_{1}^{\dagger}\left(t,s_{1}\right),
\end{align*}
in the time-convolutionless form. As a result, the exact master equation can be given by
\begin{align*}
\partial_{t}\rho & =-i\left[H_{sys},\rho\right]+\left[L,R\left(t\right)\right]+\left[R^{\dagger}\left(t\right),L^{\dagger}\right].
\end{align*}

In Fig.~\ref{fig:M2_F1}, the dynamics of the real and imaginary parts of $F_1$ are similar to that in the case I, when $\gamma = 0.5$. They all approach to fixed values in the long time limit, for $\Omega = 0$ (Blue dashed line), $1.8$ (Green dash-dotted line). But singularities in $F_1$ real part are observed when $\Omega \approx 1$ (Red solid line). But the imaginary part of $F_1$ stays at zero. In (b) and (d), When $\gamma = 5$, the dynamics of $F_1$ is Markovian and no correlated with the central frequency $\Omega$.

\section*{Result}

In this section, we look in the two classes of couplings that we have considered, and illustrate the entanglement dynamics between the electric-, optical- and mechanical- component in the system. However, for different parameters, we observe that the non-Markovian dynamics transfers from one type to another. In order to do so, we plot the entanglement dynamics versus the central frequency of environment $\Omega$ for different types of coupling. We provide a summary of all the results obtained in this section and connect the results  with the singularities observed in the coefficient functions of the $O$ operator. 

\subsubsection*{Case I}


\begin{figure}[htp]
\centering
\includegraphics[width=10 cm]{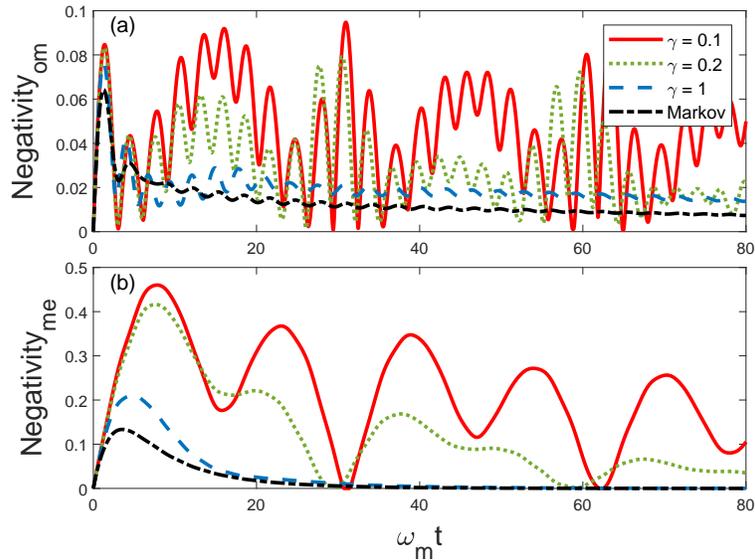}
\caption{Dynamics of entanglement between Om and me components for different decay factor $\gamma$. Other parameters are chosen as  $\omega_o = \omega_e =\omega_m$, $g_{om} = g_{me} = 0.1 \omega_m$, $\Gamma =1$.}
\label{fig:M1_Nom_101_gamma}
\end{figure}

First of all, for this exciton hopping model, we choose the initial state that there are two excitons existed in the optical cavity and the LC resonator, $\psi_{sys}(t=0) = |1\rangle _O\otimes|0\rangle_M \otimes |1\rangle_E$, to mimic the process that a photon and a phonon couple to each other via the interactions with the mechanical oscillator. Then, we investigate how does the memory time modify the entanglement dynamics. From the numerical analysis of the coefficient functions in the $\bar{O}$ operator (Fig. \ref{fig:M1_F1}), the memory factor $1/\gamma$ is dominant when it is close to the Markovian limit. We plot the dynamics of entanglement between optical cavity and mechanical oscillator ($\mathcal{N}_{om}$) in Fig. \ref{fig:M1_Nom_101_gamma}~(a), and the entanglement between mechanical oscillator and LC circuit ($\mathcal{N}_{me}$) is displayed in Fig. \ref{fig:M1_Nom_101_gamma}~(b) respectively. For the dissipative coupling described in case I, both $\mathcal{N}_{om}$ and $\mathcal{N}_{me}$ decay exponentially to zero when $\gamma =5$ and in the Markov limit. When $\gamma = 0.1$ and $0.2$, the entanglement dynamics shows the periodic behavior. And we notice that the maximal $\mathcal{N}_{om}$ happens when the $\mathcal{N}_{me}$ achieves its minimum value. This phenomena clearly display how the entanglement transfers between different components in the system and offers a potential application in realizing quantum non-demolition measurements.

\begin{figure}[htp]
\centering
\includegraphics[width=10 cm]{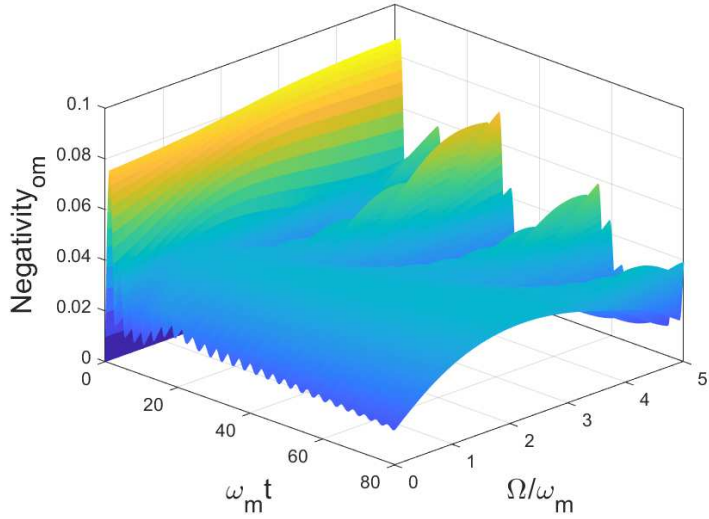}
\caption{Dynamics of entanglement between OM for different central frequencies of the environment $\Omega$. Other parameters are chosen as  $\omega_o = \omega_e =\omega_m$, $g_{om} = g_{me} = 0.1 \omega_m$, $\Gamma =1$.}
\label{fig:M1_Nom_101}
\end{figure}

\begin{figure}[htp]
\centering
\includegraphics[width=10 cm]{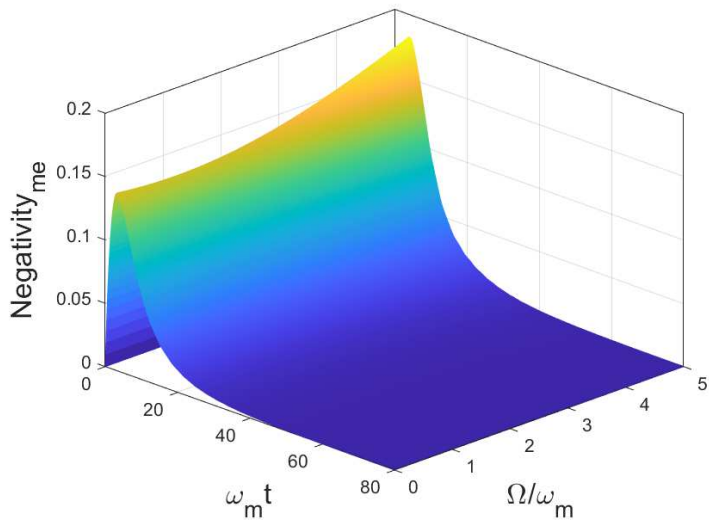}
\caption{Dynamics of entanglement between ME for different central frequencies of the environment $\Omega$. Other parameters are chosen as  $\omega_o = \omega_e =\omega_m$, $g_{om} = g_{me} = 0.1 \omega_m$, $\Gamma =1$.}
\label{fig:M1_Nme_101}
\end{figure}

In Fig. \ref{fig:M1_Nom_101}, we focus on the impact of the central frequency of the environment on the entanglement dynamics. We choose $\gamma = 1$, and adjust $\Omega$ from $0$ to $5$. In the short time interval, the generation of entanglement  $\mathcal{N}_{om}$ is caused by the linear coupling, so it is independent of the value of $\Omega$. After that, two diametrically different classes of non-Markovian dynamics are observed. The boundary of the two classes of non-Markovian dynamics locates at $\Omega \approx 1$. When $0 \leq \Omega <  1$, $\mathcal{N}_{om}$ decay non-monotonically, and after sufficient time, it decays to zero. When $\Omega > 1$, the behavior of entanglement is periodic and approaching to a non-zero value. From the above discussion, the two types of non-Markovian dynamics can be categorized by the scale of the long-time entanglement, due to the different range of central frequency of the environment $\Omega$. In Fig.~\ref{fig:M1_Nme_101}, the entanglement $\mathcal{N}_{me}$ achieve the maximal value and decay to zero quickly after for different $\Omega$ values. However, we noticed that the $\omega$ value can modify the maximal value of the negativity $\mathcal{N}_{me}$.

\subsubsection*{Case II}
\begin{figure}[htp]
\centering
\includegraphics[width=10 cm]{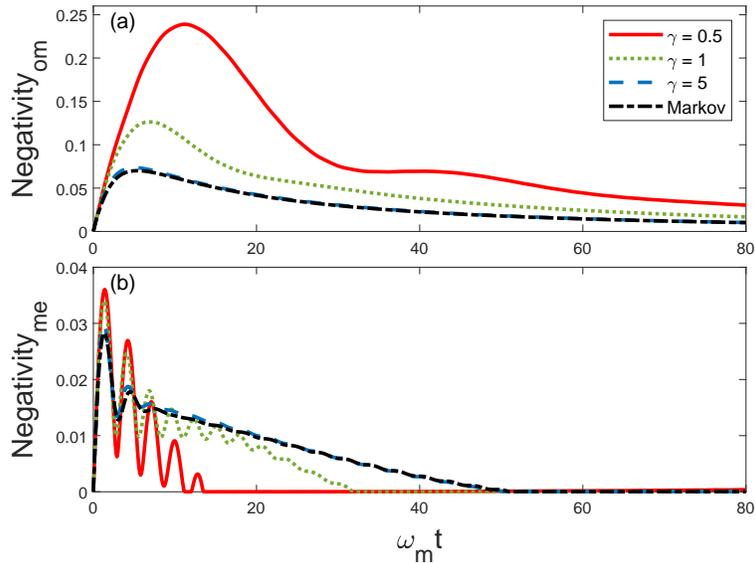}
\caption{Dynamics of entanglement between (a) OM and (b) ME, for different $\gamma$ values. Other parameters are chosen as $\omega_o = \omega_e =\omega_m$, $g_{om} = g_{me} = 0.1 \omega_m$, $\Gamma =1$. }
\label{fig:M2_Nom_000_2d}
\end{figure}
In this subsection, we consider the strong coupling model. Since the number of exciton is no longer conserved as in case I, we choose the initial state of the system as $|\psi_{sys}(t=0) = |000\rangle$, as we will focus on the generation of entanglement happened between the three components in the system. Interestingly, we find that the dynamics of $\mathcal{N}_{om}$ and $\mathcal{N}_{me}$ are very different, which can be seen in Fig. \ref{fig:M2_Nom_000_2d}. For $\mathcal{N}_{om}$ in Fig.~\ref{fig:M2_Nom_000_2d} (a), the maximal generated entanglement is positively correlated to the memory factor $1/\gamma$. While $\mathcal{N}_{me}$, in Fig.~\ref{fig:M2_Nom_000_2d} (b), entanglement sudden death (ESD) is observed for all $\gamma$ values. Moreover, the time that the ESD happens is shorter when the environment is further from the Markovian limit. At $\omega_mt 
> 60$, $\mathcal{N}_{om}$ and $\mathcal{N}_{me}$ decay to zero and indicate that the three components of the system are disentangled with each other at that instant. This feature can be used to synchronize the remote quantum network by tuning the coupling strength with the mechanical oscillator.

\begin{figure}[htp]
\centering
\includegraphics[width=10 cm]{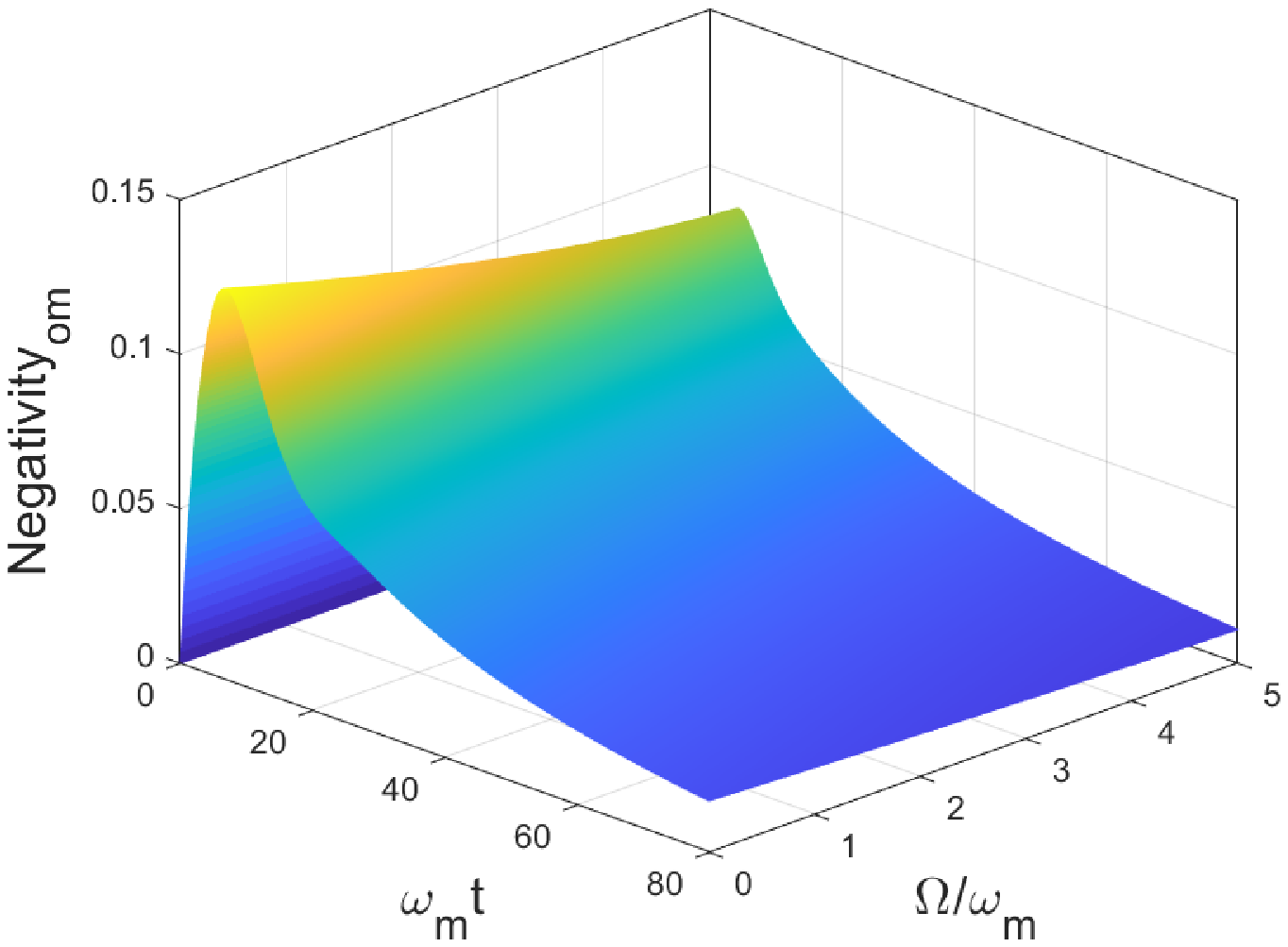}
\caption{Dynamics of entanglement between OM , for different $\omega$ values. Other parameters are chosen as $\omega_o = \omega_e =\omega_m$, $g_{om} = g_{me} = 0.1 \omega_m$, $\Gamma =1$.}
\label{fig:M2_Nom_000}
\end{figure}

\begin{figure}[htp]
\centering
\includegraphics[width=10 cm]{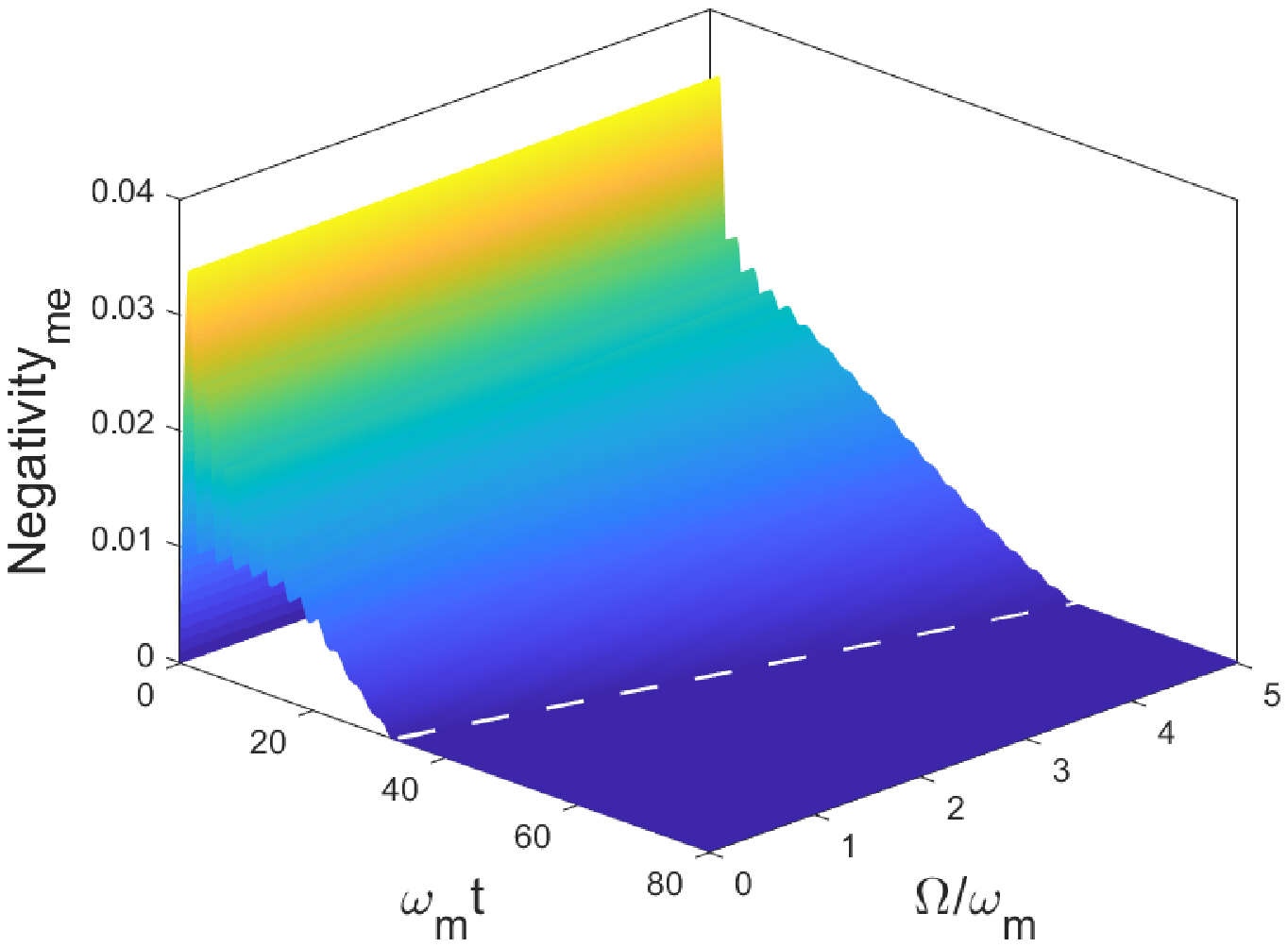}
\caption{Dynamics of entanglement between ME, for different $\Omega$ values. The dashed line marks the time that entanglement sudden death happens. It shows that the time to observe ESD has a linear relationship with the central frequency of the environment $\Omega$. Other parameters are chosen as $\omega_o = \omega_e =\omega_m$, $g_{om} = g_{me} = 0.1 \omega_m$, $\Gamma =1$.}
\label{fig:M2_Nme_000}
\end{figure}

In Fig.~\ref{fig:M2_Nom_000}, impact of the central frequency of the environment on the entanglement dynamics is that $\Omega$ can restrain the maximal generated entanglement between OM. But the long-time limit entanglement approaches to zero. In Fig.~\ref{fig:M2_Nme_000}, the generation of entanglement $\mathcal{N}_{me}$ achieves its maximal value and then decay to zero in the long-time limit. But we notice that the non-Markovian behavior can be categorized in two types again: (1), $\Omega <1$, $\mathcal{N}_{me}$ oscillates around a fixed value in the short-time regime and quickly decay to zero right after; (2), $\Omega >1$, $\mathcal{N}_{me}$ decays to zero linearly. In addition, we study the time when ESD happens ($t_{ESD}$) and find out that the time and the central frequency of the environment $\Omega$ have a quantitative relationship,
\begin{align*}
    t_{ESD} \approx \beta\Omega,
\end{align*}
which supplies a quantitatively method to estimate ESD and design the control and measurement schemes, where $\beta$ is the proportional factor and its value depends on the parameters chosen in the numerical simulations.

\section*{Conclusions}
In this paper, we study the entanglement dynamics and transfer between OM and ME components in the piezoelectric optomechanical system. The non-Markovian effects from the environment are discussed in weak coupling and strong coupling respectively. In particular, the effects of the central frequency of the environment on the entanglement dynamics are discussed. By using the quantum-state diffusion approach and the corresponding master equation, We analyze the coefficient functions evolution and discover that singularities emerge when the central frequency of the environment is around a special value. As a result, the non-Markovian dynamics of the system can be categorized in two types: (1) the entanglement gradually decays to zero; (2) the entanglement shows periodic behavior and approaches to a fixed value. This categorization can help prepare the quantum system and its coupled environment toward steady states and realize quantum teleportation in the quantum network. Moreover, the entanglement generation is studied. An quantitative conclusion is that the time to observe ESD $t_{ESD}$ is proportional to $\Omega$, which offers a robust and simple method to estimate the entanglement in the quantum network. Our work can be extended to study the non-equilibrium dynamics in a full quantum framework when the environment temperature is close to zero Kelvin. 

\bibliography{Reference}

\section*{Acknowledgements}

Y.C. gratefully acknowledges financial supports by the Institutional Support of Research and Creativity (ISRC) grant provided by New York Institute of Technology.

\section*{Author contributions statement}

Y.C. designed the study and derived the mathematical model. Q.D. and P.Z. performed numerical simulations and prepared the manuscript under the guidance of Y.C. and Y. M. All authors have read and agreed to the published version of the manuscript.

\section*{Additional information}

Competing Interests: The authors declare no competing interests.

\end{document}